\documentclass{JHEP3}
\def\bu{\bar{u}}  \def\bphi{\bar{\phi}}
  
\def\nc{\newcommand}
\nc{\pa}{\partial}  \nc{\hf}{\frac{1}{2}}         
\nc{\ben}{\begin{equation}}
\nc{\een}{\end{equation}}
\nc{\bea}{\begin{eqnarray}}
\nc{\eea}{\end{eqnarray}}
\nc{\braket}[1]{\langle\,{#1}\rangle}

\nc{\C}{\mbox{\hspace{1.24mm}\rule{0.2mm}{2.5mm}\hspace{-2.7mm} C}}
\nc{\Nat}{\mbox{\hspace{.04mm}\rule{0.2mm}{2.8mm}\hspace{-1.5mm} N}}
\nc{\spa}{\hspace{1 cm},\hspace{1 cm}}
\nc{\vs}{\vspace}
\nc{\NP}[1]{Nucl.\ Phys.\ {\bf #1}}
\nc{\PL}[1]{Phys.\ Lett.\ {\bf #1}}
\nc{\CMP}[1]{Commun.\ Math.\ Phys.\ {\bf #1}}
\nc{\LMP}[1]{Lett.\ Math.\ Phys.\ {\bf #1}}
\nc{\TMP}[1]{Theor.\ Math.\ Phys.\  {\bf #1}}
\nc{\PR}[1]{Phys.\ Rev.\ {\bf #1}}
\nc{\PRL}[1]{Phys.\ Rev.\ Lett.\ {\bf #1}}
\nc{\PTP}[1]{Prog.\ Theor.\ Phys.\ {\bf #1}}
\nc{\PTPS}[1]{Prog.\ Theor.\ Phys.\ Suppl.\ {\bf #1}}
\nc{\MPL}[1]{Mod.\ Phys.\ Lett.\ {\bf #1}}
\nc{\IJMP}[1]{Int.\ Jour.\ Mod.\ Phys.\ {\bf #1}}
\nc{\IM}[1]{Invent.\ Math.\ {\bf #1}}
\nc{\SJNP}[1]{Sov. J. Nucl. Phys.\ {\bf #1}}

\def\ni{\noindent}
\def\wt{\widetilde}      
      \def\ol{\overline}
\def\nn{\nonumber \\}

\def\sc#1{{\scriptstyle #1}}

\def\to{\rightarrow}

            \def\F{{\cal F}}

\def\Z{{\bf Z}}
\def\R{{\bf R}}
\def\C{{\bf C}}

\def\bz{{\bar z}}      \def\bw{{\bar w}}

\usepackage[dvips]{graphicx}

\author{
Yukitaka Ishimoto \\
Theoretical Physics Laboratory, RIKEN, Saitama 351-0198, JAPAN\\
E-mail: y.ishimoto@riken.jp
}
\title{Two-Point Functions and Logarithmic Boundary Operators 
in Boundary Logarithmic Conformal Field Theories}
\abstract{
Amongst conformal field theories,
there exist logarithmic conformal field theories 
such as $c_{p,1}$ models. 
We have investigated $c_{p,q}$ models with a boundary in search of logarithmic
theories and have found logarithmic solutions of two-point functions in the context of the Coulomb gas picture. We have also found the relations between coefficients in the two-point functions and correlation functions of logarithmic boundary operators, and have confirmed the solutions in \cite{KW}. Other two-point functions and boundary operators have also been studied in the free boson construction of boundary CFT with $SU(2)_k$ symmetry in regard to logarithmic theories. This paper is based on a part of D. Phil. Thesis \cite{Ishimoto:2003nb}.
}
\preprint{RIKEN-TH-25}
\dedicated{Dedicated to the memory of Ian I. Kogan}
\begin{document}
\section{Introduction}
\label{ch:intro}

Since \cite{BPZ}, 
there is no doubt that 2d conformal field theory (CFT) has accomplished a beautiful success in a variety of physics. Among them, there is a class of theories called {\it Logarithmic CFT} which contains degenerate fields, or logarithmic pairs $C$ and $D$ in its simplest case \cite{gura}. A definition of the pair can be given by their operator product expansions and correlation functions \cite{gura,CKT}:
\begin{eqnarray}
 \label{def:CD}
 \label{def:rank2JS}
  T(z) C(w) &\sim& \frac{h_C \,C(w)}{(z-w)^2} + \frac{\partial_w C(w)}{z-w} , \nn
  T(z) D(w) &\sim& \frac{h_C \,D(w) + C(w)}{(z-w)^2} + \frac{\partial_w D(w)}{z-w} ,
  \nn
  \braket{C(z) C(w)} &\sim& 0 , \quad
  \braket{C(z) D(w)} \sim \frac{\alpha}{(z-w)^{2 h_C }} , \nn
  \braket{D(z) D(w)} &\sim& \frac{1}{(z-w)^{2 h_C }}\left(-2 \alpha \ln (z-w) + \alpha^{\prime} \right) ,
\end{eqnarray}
where $T(z)$ is the stress tensor and $h_C$ is the conformal dimension of the pair. From the above, one can read off that the pair forms a $rank$-2 Jordan cell, or a reducible but indecomposable representation under the Virasoro zero mode of $T(z)$.
There may be different ways of finding such structure, but logarithmic singularities in four-point functions can be interpreted as its origin, or the essential evidence of it \cite{Knizhnik:xp}.
For various models and preliminary knowledge of logarithmic CFT, see for example \cite{Ishimoto:2003nb} and references therein.

The study of logarithmic CFT is still in progress though, much progress has been made so far.
Despite this, only a small number of papers have contributed to logarithmic CFT with boundaries (boundary LCFT) 
\cite{Ishimoto:2003nb,KW,ishimoto1,YI2} 
in the context of boundary CFT initiated by Cardy \cite{Cardy:bb,card1,Cardy:tv}.
In this untamed conformal zoo, the first systematic attempt to formulate a boundary LCFT and boundary states was made by Kogan and Wheater in \cite{KW}, where they showed some boundary two-point functions and their relations with logarithmic boundary operators in a few models, being based solely on the analysis of the differential equations.

Our main aim in this paper is to explicitly compute such boundary two-point functions and to find their relations with {\it logarithmic} boundary operators in order to show a way to more complicated correlators and relations in boundary LCFT. In addition to the above, another focus is on the conditions for the emergence of logarithmic CFT: what models may be logarithmic CFT? The latter can be found during the former, because it is almost equivalent to see the conditions for logarithmic singularities in the correlation functions. The former requires some computational effort and an explicit field representation which has been developed in a firm basis.

Among free field representations, the Coulomb gas picture (CG), or
Dotsenko-Fateev construction, is a powerful computational tool and is known to represent a wide variety
of CFTs in two dimensions \cite{Dotsenko:1984nm,Dotsenko:1984ad}.
Besides, free boson realisations are available and useful in various
models such as the $SU(2)_k$ Wess-Zumino-Novikov-Witten models (WZNW models \cite{WZNW}) via bosonisation \cite{Gerasimov:fi}.
On the other hand, boundary CFT of such representations has only recently been studied
in a comparatively small number of papers \cite{Ishimoto:2003nb,schulze,kawaic1,yi talk,Caldeira:2003zz,Hemming:2004dm} and logarithmic CFT with no boundary in the CG picture \cite{Hata:2000zg,Nichols:2003dj}, while any of them is not for boundary LCFTs (except \cite{Ishimoto:2003nb}).

Another motivation to study such models is that there appears a close relation of logarithmic CFT to some geometrical interpretations of the string backgrounds (see for example \cite{Rasmussen:2004vx}). Together with the existing interests of D-brane physics \cite{stringy}, 
boundary logarithmic CFT seems to enter into the games.

In section \ref{ch:BCFT}, we give a short introduction to boundary CFT on an upper half-plane, following  Cardy's works.
In the following section \ref{sec:CG}, we apply boundary CFT and the CG picture to the $c_{p,q}$ model on the upper half plane, mapping it to a chiral theory on the whole complex plane. 
Following CG techniques for four-point functions, we obtain the boundary two-point functions which exhibit logarithmic singularities. In this type of models, two conditions for logarithms (logarithmic conditions) will be given in search of LCFTs. Thereafter, the relations of the two-point functions and the normalisation factors of the logarithmic boundary operators, are deduced, by which we confirm the results in \cite{KW}.

It was shown that the $SU(2)_k$ WZNW models can be described by free field realisations and, in particular, by three free bosons by Gerasimov et al. \cite{Gerasimov:fi}.
In addition, some of the $SU(2)_k$ WZNW models have been shown, or suggested, to be logarithmic CFT \cite{Gaberdiel:2001ny,Nichols:2003gm,Caux:1996kq,nichols1}. 
In section \ref{sec:su2}, we give boundary two-point functions of the doublet representations in the boson realisation defined on the upper half plane. Confirming the consistency with other results such as Knizhnik-Zamolodchikov equations (KZ-equation) \cite{Knizhnik:nr}, we obtain logarithmic conditions which contain $k=0$ case and study the relations between the two-point functions and the logarithmic boundary operators as in section \ref{sec:CG} for $k=0$.

This paper is based on a part of D. Phil Thesis \cite{Ishimoto:2003nb} with some new discussions on the logarithmic conditions in conclusions.

\subsection{Boundary CFT: on an Upper Half-Plane}
\label{ch:BCFT}
\label{sec:CFT plane}

\begin{figure}[h]
\centering
\includegraphics{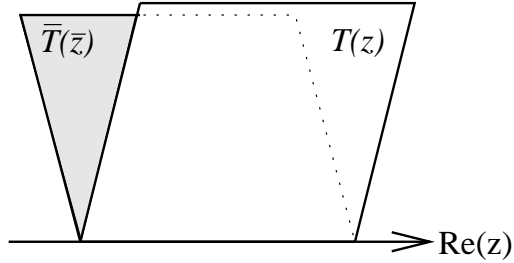}
 \caption{Holomorphic and anti-holomorphic sectors defined on a UHP
 are equivalent to one sector defined on two UHPs but connected to
 each other at the boundary.}
      \label{fig:2UHP}
\end{figure}

Boundary CFT is defined on a two-dimensional surface with
one or more boundaries and the prototype of its geometry is an upper
half-plane (UHP). On a UHP, the theory is decomposed into and described by holomorphic and anti-holomorphic sectors, 
however they are not totally independent but coupled to
each other by the boundary conditions.
In this case the boundary is the real axis (Fig.\ref{fig:2UHP}). 

\ni
The
boundary itself must be invariant under conformal transformations.
The allowed infinitesimal conformal coordinate transformations are $z\to z+ \epsilon(z)$ such that
$
 \label{def:conformally invariant bc}
  \epsilon(z) \in \R ~{\rm if~ z\in\R} .
$
This makes it a real analytic function $\epsilon(\bz)=\ol{\epsilon(z)}$.
On operators, the conformal transformations are expressed 
by a contour integral:
\begin{eqnarray}
 \label{def:BCFTtr}
   \delta_\epsilon = \oint dz\, \epsilon(z) T(z) - \oint d\bz\, \epsilon(\bz) \ol T(\bz) , 
\end{eqnarray}
where the closed contours surround all the relevant operators on the UHPs respectively. This actually produces the conformal Ward identity on correlation functions. For the identity to be valid, the above integral on the boundary should vanish and it gives the following condition
\bea
 \label{def:bc_T}
   T(x)=\ol T(x), \ x\in\R ,
\eea
for this case. This is equivalent to $T_{tx}=0$ in the Cartesian coordinate, meaning no energy flows across the boundary. 
In (\ref{def:BCFTtr}), one can take the contours so that two infinite semi-circles are tied up at the boundary without contours on the axis and, if we map the anti-holomorphic sector to the LHP, the theory can be interpreted as a chiral theory defined on the whole plane.

$N$-point functions in boundary CFT are now re-interpreted as $2N$-point functions on the whole plane in a chiral theory and therefore the differential equations for four-point functions are also naturally defined for the minimal models. 
For example, a one-point function of the field $\Phi_\Delta$ with conformal dimension $\Delta$ can be drawn as
\bea
 \label{def:2pt anonymous}
   \braket{\Phi_\Delta(z,\bz)}_{boundary}
   = \braket{\Phi_\Delta(z) \Phi_\Delta(\bz)}
   = (z-\bz)^{-2 \Delta}.
\eea
The final equivalence is exact up to phase.
Since $(z+\bz)\in \R$, operators appearing in the $r.h.s.$ of their operator product expansion become special having no anti-holomorphic counterparts: $\Phi_\Delta(z) \Phi_\Delta(\bz) \sim \sum_{\Delta_c} \Phi_{\Delta_c}(\frac{z+\bz}{2} \in \R)$. This special type of operator $\Phi_{\Delta_c}(x \in \R)$ is called `boundary operator' and is important in the construction of boundary CFT.

\section{The Coulomb Gas Picture of LCFT with Boundary}
\label{sec:CG}

In this section, we consider the Coulomb gas picture of the $c_{p,q}$ models \cite{BPZ,Dotsenko:1984nm,Dotsenko:1984ad} with a boundary, for LCFTs in particular. 
We will later focus on a $c_{2,1}=-2$ case with the Neumann boundary condition.

\subsection{The $c_{p,q}$ Models with Boundary}
\label{sec:pq}

The action functional of a free boson $\Phi$ is given by:
\begin{eqnarray}
\label{def:action pq-model}
 S_{CG}[\Phi] = \frac1{8\pi} \int_\Sigma d^2 z \sqrt{g} 
 \left[ g^{\mu\nu} \pa_\mu \Phi(z,\bz) \pa_\nu \Phi(z,\bz) + 2\,i \alpha_0 R^{(2)} \Phi(z,\bz) \right].
\end{eqnarray}
$\alpha_0$ is the background charge $\alpha_0 = (p-q)/\sqrt{2pq}$,
and the central charge is $c_{p,q}=1-12\alpha_0^2=1-\frac{6(p-q)^2}{pq}$ with
$p,q\not=0$. 
$R^{(2)}$ is the two-dimensional scalar curvature of the 2d Riemann surface $\Sigma$ with a boundary. We set $\Sigma$ to be the upper half plane (UHP), $g_{\mu\nu}=\left(\begin{array}{cc} 
			       0  & 1/2 \\
			       1/2 & 0 
				    \end{array} \right)$, 
and assume $p=1$ or $q=1$ or $(p, q)$ coprime integers. Note, however, that we can formally consider non-coprime cases in regard to continuous flows of $p$ and $q$. Another important restriction on them will come up later in (\ref{assum:pq}).

The equation of motion (EOM) implies that $\Phi(z,\bz)$ is analytic in the bulk, that is, $\Phi(z,\bz)= \phi(z) + \bphi(\bz)$ with their propagators:
\begin{eqnarray}
 \braket{\phi(z)\phi(w)} = -\ln (z-w)\,,\,\,  \braket{\bphi(\bz)\bphi(\bw)} = -\ln (\bz-\bw) .
\end{eqnarray}
The stress tensor
$T(z)$ (the energy-momentum tensor) is given by:
\begin{eqnarray}
 T(z) = - \frac{4\pi}{\sqrt{g}} \frac{\delta S[\phi]}{\delta g^{zz}} = - \frac12 :\left(\pa\phi(z)\right)^2: + i \alpha_0 \pa^2 \phi(z) ,
\end{eqnarray}
while its anti-holomorphic counterpart is given similarly.
The normal ordering ($: operators :$) is introduced to the first term to
subtract its divergence.

There are two different boundary conditions, but we only deal with the Neumann boundary condition in what follows.\footnote{
The Dirichlet boundary condition in CG was discussed in \cite{schulze}.
}
In either case, the holomorphic part and the anti-holomorphic part can no longer be independent of each other: $\braket{\phi\bphi}\ne 0$ for general $z,w$.
Indeed, by the ``method of images'' initiated by Cardy \cite{Cardy:bb}, the
anti-holomorphic sector of the theory can be mapped onto the lower
half-plane ($LHP$) by $\bphi(\bz) = \Omega \phi(z^*)$ with the parity 
$\Omega=1$ ($\Omega=-1$ for the Dirichlet case). In the given geometry,
$\bz=z^*$, so $\bphi(\bz) = \Omega \phi(\bz=z^*)$. 
Thereby, all the propagators
$\braket{\phi\phi}$, $\braket{\bphi\bphi}$, $\braket{\phi\bphi}$, $\braket{\bphi\phi}$,
are simplified to $\braket{\phi\phi}$ with corresponding arguments, 
and the theory becomes effectively chiral. Also, Cardy's equation on
the boundary:
\begin{eqnarray}
 T(z)=\ol T(\bz) \Bigr|_{\pa\Sigma}
\end{eqnarray}
is trivially satisfied, and the conformal transformations of the primary
fields are generated by the single tensor field $T(z)$ being analytically
continued to $z\in\C$.

The chiral primary Kac fields $\Phi_{r,s}(z)$ of the $c_{p,q}$ 
model have a pair of integer indices $r,s$.
Their conformal dimensions obey the Kac formula \cite{kac}: $h_{r,s}=
-\frac12 \alpha_0^2 + \frac18 \left( r \alpha_+ + s \alpha_- \right)^2$ 
with $\alpha_\pm \equiv \alpha_0 \pm \sqrt{\alpha_0^2 + 2}$.
Assuming our non-chiral primaries are $ \Phi_{r,s}(z,\bz) = \Phi_{r,s}(z) \otimes \bar\Phi_{r,s}(\bz) $, the method of images implies that:
\begin{eqnarray}
\Phi_{r,s}(z,\bz) = \Phi_{r,s}(z)\,\Phi_{r,s}(z^*)
\end{eqnarray}
Then, a problem of a boundary $N$-point function reduces 
to that of a chiral $2N$-point function. 
For example:
\begin{eqnarray}
\label{def:2pt-4pt}
 \braket{\Phi_{r,s}(z,\bz)\Phi_{r^\prime,s^\prime}(w,\bw)}_{boundary} 
= \braket{\Phi_{r,s}(z)\Phi_{r^\prime,s^\prime}(w)\Phi_{r^\prime,s^\prime}(w^*)\Phi_{r,s}(z^*)}_0 \,\,.
\end{eqnarray}
$\braket{\cdots}_0$ denotes the vacuum expectation value in the mapped
chiral theory. Its charge condition in CG will appear shortly.

The vertex operator realisation provides two ways
of representing the chiral primaries due to the presence of conjugate operators:
\begin{eqnarray}
\label{def:V_{r,s}}
 \Phi_{r,s}(z) \sim V_{r,s}(z) = V_{\alpha_{r,s}} = :e^{i \alpha_{r,s} \phi}:{\rm ~~or~~} \wt V_{r,s}(z) = V_{2\alpha_0 - \alpha_{r,s}} .
\end{eqnarray}
The charges are $\alpha_{r,s}=\frac{1-r}2 \alpha_+ + \frac{1-s}2 \alpha_-$.
The screening currents of dimension $(h,\bar h) = \{(1,0), (0,1)\}$ are defined as:
\begin{eqnarray}
\label{def:scop BCFT}
 Q_\pm \equiv \oint_{C} d\zeta\, J_\pm (\zeta), \; J_\pm \equiv V_{\alpha_\pm},
\end{eqnarray}
with the contour $C$ on $\Sigma$. 
The neutrality condition (charge asymmetry condition
in \cite{Dotsenko:1984nm}) of $\braket{V_{\alpha_1}(z_1) \cdots V_{\alpha_N}(z_N)\, Q_+^{\,n_+} Q_-^{\,n_-} \ol Q_+^{\,\bar n_+} \ol Q_-^{\,\bar n_-}}_0$ is thus: 
\begin{eqnarray}
\label{def:charge cond pq}
  \sum_{i=1}^N \alpha_i + \sum_{\epsilon=+,-} 
  \left(n_\epsilon + \bar n_\epsilon\right) \alpha_\epsilon
 = 2 \alpha_0 \;\;. 
\end{eqnarray}
Note that $p=q$ is a special point where $\alpha_0$ becomes
zero and the screening charges are not required, as $\braket{V_{\alpha_1}\wt V_{\alpha_1}\cdots V_{\alpha_N}\wt V_{\alpha_N}}$ satisfies the condition (\ref{def:charge cond pq}).

\subsection{Boundary Two-Point Functions of $\Phi_{1,2}$ and $\Phi_{r,s}$ Fields with Logarithms}

Suppose $p\not= q$.
Consider a boundary two-point function with the Neumann
boundary condition:
$
 \braket{\Phi_{r,s}(z_1,\bz_1) \Phi_{1,2}(z_2,\bz_2)}_{boundary} \,.
$
From eq. (\ref{def:2pt-4pt}), this can be expressed in a chiral and
symmetric form, and 
with (\ref{def:V_{r,s}}, \ref{def:charge cond pq}), there are two choices in its vertex operator realisation. 
The general form of the two-point function is given by a linear combination of them:
\begin{eqnarray}
\label{eq:linear comb pq}
\braket{\Phi_{r,s}(z_1,\bz_1) \Phi_{1,2}(z_2,\bz_2)}_{boundary}
&=& \braket{\Phi_{r,s}(z_1) \Phi_{1,2}(z_2) \Phi_{1,2}(z_2^*) \Phi_{r,s}(z_1^*)}_0
\nn
&=& \alpha_1^{(pq)}\, F_1(z_1, z_2) + \alpha_2^{(pq)}\, F_2(z_1, z_2) \,,
\end{eqnarray}
where $\alpha_1^{(pq)}, \alpha_2^{(pq)}$ are, so far, arbitrary constants 
and the two choices are:
\begin{eqnarray}
\label{def:2 int pq}
F_1(z_1, z_2) &\equiv& \braket{V_{r,s}(z_1) V_{1,2}(z_2) V_{1,2}(z_2^*) \wt V_{r,s}(z_1^*)\; Q_- } \,,
\nn
F_2(z_1, z_2) &\equiv& \braket{V_{r,s}(z_1) V_{1,2}(z_2) V_{1,2}(z_2^*) \wt V_{r,s}(z_1^*)\; \ol Q_- } \,.
\end{eqnarray}

\begin{figure}[h]
\centering
\includegraphics[scale=0.5]{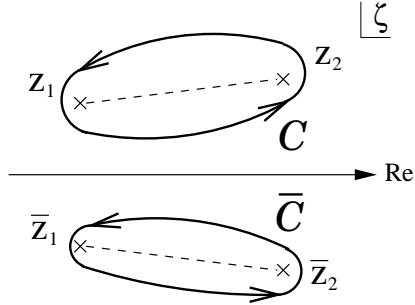}
 \caption{Non-contractable contours $C,\,\bar C$ for the $p=2$
 case. Dotted lines are branch cuts.}
      \label{fig:pochhammer}
\end{figure}

\begin{figure}[h]
\centering
\includegraphics[scale=0.55]{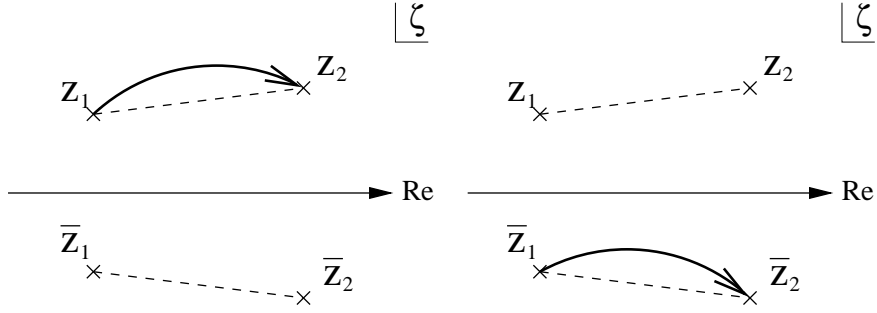}
 \caption{The line integrals from $z_1$ to $z_2$ (Left) and from $\bz_1$ to $\bz_2$ (Right).}
      \label{fig:line int}
\end{figure}

The problem of $\braket{\Phi_{r,s}\Phi_{1,2}}_{boundary}$ is now 
reduced to the calculation of two integrals.
First we have:
\begin{eqnarray}
\label{eq:rs-12 vertex}
F_1 &=& \oint_{C\subset\Sigma} \!\!\!d\zeta\, 
 \braket{V_{r,s}(z_1) V_{1,2}(z_2) V_{1,2}(z_2^*) \wt V_{r,s}(z_1^*)\,J_-(\zeta) }
\nn
&=& \prod_{i<j} {z_{ij}}^{\alpha_{i} \alpha_{j}} \oint_{C\subset\Sigma} \!\!\!d\zeta\, 
 \prod_{i=1}^4 \left( z_i - \zeta \right)^{\alpha_{i} \alpha_-} ,
\end{eqnarray}
with $\alpha_2=\alpha_3=\alpha_{1,2}$, $\alpha_1=\alpha_{r,s}$,
$\alpha_4= 2 \alpha_0 - \alpha_{r,s}$,
and $z_{ij}=z_i - z_j$, $z_3=z_2^*$, $z_4=z_1^*$.
$F_2$ can be written similarly.
Since the exponents in the integrand are:
\begin{eqnarray}
\label{eq:exponents}
&& \alpha_1 \alpha_- = 
\frac1p \left( (1-s)q-(1-r)p \right),\,\, 
 \alpha_4 \alpha_- = \frac1p \left( (1+s)q-(1+r)p \right) , 
\nn
&&
\alpha_2 \alpha_- = \alpha_3 \alpha_- 
 = - \frac{q}p \,\,, 
\end{eqnarray}
$\zeta= z_1, z_2, z_3, z_4$ are branch points of order $p$, 
when $p,q$ are coprime integers and $(1\pm s)$ are not multiples of $p$.
Nontrivial non-contractable contours $C,\bar C$ can be reduced to the ones encircling two points, or the Pochhammer-type contour between the two points, on each half-planes (see Fig.\ref{fig:pochhammer} for $p=2$).
Assuming that:
\begin{eqnarray}
\label{assum:pq}
 \frac{q}{p}\not\in\Z {~~~and~~~} \frac{q\,(1\pm s)}{p}\not\in\Z ,
\end{eqnarray}
both closed contour integrals can be further reduced to two line integrals of 
$z_1\to z_2$ and of $z_1^* \to z_2^*$ up to constant 
(Fig.\ref{fig:line int}). For $(p,q)$ coprime cases, the first condition is automatically satisfied.

By a standard choice of conformal mappings $u=\frac{z_{34}}{z_{31}}\frac{\zeta-z_1}{\zeta-z_4}$ : $(z_1,z_2,z_3,z_4)\to (0,\xi,1,\infty)$, 
the interval $(z_1,z_2)$ is mapped to $C_1=(0,\xi)$ and the integral is transformed into:
\begin{eqnarray}
\label{eq:F1 pq}
 F_1 = 
 z_{14}^{-2 h_{r,s}} z_{23}^{-2 h_{1,2}} \xi^{\alpha_1 \alpha_2} \left(1-\xi\right)^{2\alpha_2 (\alpha_2 - \alpha_0)}
 \int_0^\xi du\, u^{\alpha_1 \alpha_-} (\xi-u)^{\alpha_{2} \alpha_-}  (1-u)^{\alpha_{2} \alpha_-} ,
\end{eqnarray}
up to phase.
$\xi$ is an harmonic ratio, 
$\xi=\frac{z_{12}z_{34}}{z_{13}z_{24}}=\left|\frac{z_1-z_2}{z_1-z_2^*}\right|^2$
and $0<\xi<1$.
$F_2$ is given similarly by the contour $C_2=(1,\infty)$.

The integrals in $F_i$'s are the integral representations of the first two
Kummer's solutions of the hypergeometric differential equation \cite{slater}:
\begin{eqnarray}
\label{gauss eq}
 \left[ \xi (1- \xi) \frac{d^2}{d\xi^2} 
 + \left\{c - \left(1+a+b\right)\xi \right\} \frac{d}{d\xi} -a b \right]
 F_i (\xi) = 0 \,,
\nn
\left\{
\begin{array}{rcl}
 a =& \alpha_-^2/2 &= \frac{q}{p} \\
 b =& -1 - \left(\alpha_1 - \alpha_- \right)\alpha_- 
    &= \frac{q}p \left(1+s\right) - r \\
 c =& - \left( \alpha_1 -\alpha_-/2 \right) \alpha_- 
    &= \frac{q}p\;s -r +1 
\end{array}
\right. \,,
\end{eqnarray}
\begin{eqnarray}
\label{sol:pq}
 F_1 &\sim& \frac{\Gamma(1+a-c)\Gamma(1-a)}{\Gamma(2-c)}\; \xi^{1-c}\; 
  {}_2F_1\left(1+a-c,1+b-c;2-c;\xi \right) \,, 
\nn
 F_2 &\sim& \frac{\Gamma(b)\Gamma(c-b)}{\Gamma(c)}{}_2F_1\left(a,b;c;\xi \right) \,.
\end{eqnarray}
The common prefactor of the integral in (\ref{eq:F1 pq}) is neglected. 
The form of the solutions (\ref{sol:pq}) is
essentially same as $\braket{\phi_{n,m}\phi_{1,2}\phi_{1,2}\phi_{n,m}}$
in \cite{Dotsenko:1984nm}.
This observation also agrees with 
$\braket{\phi_{1,2}\phi_{1,2}}_{boundary}$ in \cite{kawaic1}.
It should be mentioned that eq. (\ref{gauss eq}) is in complete
agreement with the differential equation that the function
(\ref{eq:linear comb pq}) of the Kac fields must satisfy
\cite{BPZ,Dotsenko:1984nm,Kogan:1997fd,KW}. 
Thus, the linear combination in eq. (\ref{eq:linear comb pq}) indeed
gives the general solution.

However, 
as was discussed in \cite{gura,Kogan:1997fd}, the second-order linear differential
equation may possess a solution logarithmic at $\xi=0$, through the Frobenius
method or other similar methods, and lead to a logarithmic field of LCFT.
In such {\it logarithmic} cases, either two hypergeometric functions in the solutions (\ref{sol:pq}) coincide, or one of them becomes indefinite at $\xi=0$. 
They do not provide two independent solutions of
eq. (\ref{gauss eq}). 
The necessary condition for such an emergence of logarithms is 
that the coefficient $c$ takes integer value \cite{slater}.
Since $r$ and $s$ are integers by definition, the condition can be translated
into that of $p,q$, and $s$:
\begin{eqnarray}
\label{cond:log}
 s\cdot \frac{q}{p} \in \Z ,
\end{eqnarray}
which is compatible with the condition (\ref{assum:pq}).

Under the conditions (\ref{assum:pq}, \ref{cond:log}), one can explicitly
observe that the two solutions (\ref{sol:pq}) become identical:
$F_1 = F_2$, up to phase.
Consequently, one can write down the solution for integer $c$ as:
\begin{eqnarray}
\label{sol:pq2}
 \braket{\Phi_{r,s}(z_1,\bz_1)\Phi_{1,2}(z_2,\bz_2)}_{boundary}
= \alpha \;C(z_1,z_2)\; 
{}_2 F_1\left(\frac{q}{p}, \frac{q}{p} +|n| ; 1+|n|; \xi\right) \,,\,\,
\end{eqnarray}
where $n=(\frac{q}{p} s-r)\in\Z$, $\alpha$ is an arbitrary constant.
The function $C(z_1,z_2)$ is the prefactor in eq. (\ref{eq:F1 pq}):
\begin{eqnarray}
 C(z_1,z_2) = 
(z_1-\bz_1)^{-2 h_{r,s}} (z_2-\bz_2)^{-2 h_{1,2}} 
 \xi^{\frac1{2p}\left\{(s-1)q+(1-r)p\right\}} \left(1-\xi\right)^{\frac{2q}{p} -1} ,
\end{eqnarray}
with the conformal dimensions 
$h_{r,s}=\frac12 \alpha_{r,s}(\alpha_{r,s}- 2\alpha_0)$ and 
$h_{1,2}= \frac12 \alpha_{1,2}(\alpha_{1,2}- 2\alpha_0)$.
The theory of hypergeometric functions further shows that the
above solution is logarithmic at $\xi=1$ in some cases. In 
section \ref{sec:c=-2 bop}, we will see this logarithm and discuss its meaning in a
particular case, $c_{2,1}=-2$.

\subsection{Another Operator for the Logarithmic Solutions}
\label{sec:logsol}

Define two screening currents analytically continued from $\Sigma$ to $\C$:
\begin{eqnarray}
\label{def:new screener}
  \wt J_\pm = V_{\alpha_\pm} (z) ~~~for~~z\in\C.
\end{eqnarray}
Recall that the conformal transformations are realised by the single
tensor field $T$, and the conformal dimension of $\wt J_\pm$ is 
$\wt h=1$ under this $T$. 
By making a closed contour integration on the whole
 plane, one can define the analytically continued version of
the screening charges:
\begin{eqnarray}
\label{def:new charge}
 \wt Q_\pm = \oint_{\wt C} d\zeta \wt J_\pm(\zeta) . 
\end{eqnarray}
Here, $\wt C$ can lie anywhere on the whole complex plane $\C$.

\begin{figure}[h]
\centering
\includegraphics[scale=0.55]{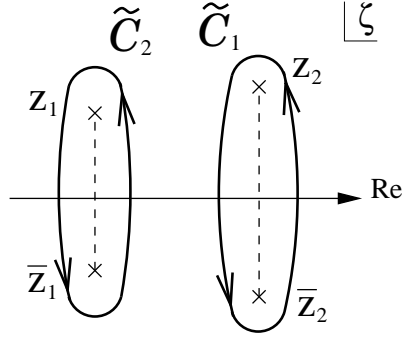}
 \caption{Two non-contractable contours for $\wt C$ for the $p=2$
 case. Subscripts of the contours are for later convenience.}
      \label{fig:c23}
\end{figure}

With these operators, one can compute a chiral four-point function:
\begin{eqnarray}
 \wt F(z_1,z_2) &=& \braket{V_{r,s}(z_1) V_{1,2}(z_2) V_{1,2}(z_2^*) V_{r,s}(z_1^*) \,\wt Q_-} . 
\nn
 &=& \prod_{i<j} z_{ij}^{\alpha_i \alpha_j} \oint_{\wt C\subset \C} d\zeta \prod_{i=1}^4 \left(z_i-\zeta\right)^{\alpha_i \alpha_-} ,
\end{eqnarray}
with the exponents (\ref{eq:exponents}).
The only difference between the above and eq. (\ref{eq:rs-12 vertex}) is
 the contour $\wt C$, so we restrict $\wt C$ to be analytically
 inequivalent to $C$ and $\bar C$ in eq. (\ref{eq:rs-12 vertex}) 
for nontrivial results.
The branching structure of the integrand is not unique 
(see Fig.\ref{fig:pochhammer}).
For example, a different structure with different non-contractable
contours for the branch covering of $\C$ is shown in Fig.\ref{fig:c23} for $p=2$. 
The new contours are now encircling the points $z_2$, $\bz_2$ ($z_1$, $\bz_1$), or Pochhammer-type contours between them.
The closed contours can be reduced to the line
integrals under the condition (\ref{assum:pq}).

By the same conformal mapping as for eq. (\ref{eq:F1 pq}), one obtains
for $\wt C_1$:
\begin{eqnarray}
\label{sol:wt F1}
  \wt F_1 
  &=& C(z_1, z_2)\, (1-\xi)^{1-\frac{2 q}{p}} \frac{\Gamma^2\left(1-\frac{q}{p}\right)}{\Gamma\left(2\left(1-\frac{q}{p}\right)\right)} \,
  {}_2 F_1 \left( 1-\frac{q}{p}+n, 1-\frac{q}{p}; 2\left(1-\frac{q}{p}\right); 1-\xi\right) .
\nn
\end{eqnarray}
Similarly, the line integral of $\wt C_2$ reduces to:
\begin{eqnarray}
\label{sol:wt F2}
 \wt F_2 
 &=& C(z_1, z_2)\, \frac{\Gamma^2\left(\frac{q}{p}\right)}{\Gamma\left(\frac{2q}{p}\right)}\,
  {}_2 F_1 \left( \frac{q}{p}+n , \frac{q}{p}; \frac{2q}{p}; 1-\xi\right) ,
\end{eqnarray}
where $n=\frac{q}p s-r \in\Z$.
That is to say, $\wt F_1$ and $\wt F_2$ are the sixth and
fifth Kummer's solutions of the same hypergeometric differential equation (\ref{gauss eq}),
regular at $\xi=1$.
Under the condition (\ref{assum:pq}), they both have logarithmic singularities in the vicinity of $\xi=0$, and in general they are two independent solutions in the region of $0<\xi<1$.

There is, of course, a logarithmic condition for this case:
$  2\,\frac{q}p\in\Z $.
Due to (\ref{assum:pq}), it reads:
\begin{eqnarray}
\label{cond:log2}
 \frac{q}p = \frac12 + m \,,~~ m\in\Z .
\end{eqnarray}
As in the former case, we have the following coincident solution:
\begin{eqnarray}
 \wt F = C(z_1, z_2)\, 
  \frac{\Gamma^2\left(\frac12 + |m|\right)}{\Gamma(1+2|m|)}
  {}_2 F_1 \left( \frac12+|m|+n, \frac12+|m|; 1+2|m|; 1-\xi \right) .
\end{eqnarray}
It can be checked that the function is singular at $\xi=0$ for $n\in\Z$, so this is therefore the second independent solution of
eq. (\ref{gauss eq}).
A general solution of (\ref{eq:linear comb pq}) can be given by:
\begin{eqnarray}
\label{sol:pq double}
   \braket{\Phi_{r,s}(z_1,\bz_1) \Phi_{1,2}(z_2,\bz_2)}_{boundary}
 =  \alpha F(z_1,z_2) + \wt \alpha \wt F(z_1,z_2) .
\end{eqnarray}

It should be mentioned here that, if one replaces $\wt V_{r,s}$ with
the puncture-type operator $:\phi\; \wt V_{r,s}:$ in eq. (\ref{def:2 int
pq}), the imposition of $\braket{:\phi\; V_{2\alpha_0}:} \sim 1 $ leads
to the same results appearing in this and the previous sections. 
Note again that when $p=q$ the above discussions and solutions make no
sense because screening charges are not required.

\subsection{An Example at $c_{2,1}=-2$ and the Boundary Operators}
\label{sec:c=-2 bop}

Consider the $c_{2,1}$ model in CG. The central charge is $c_{2,1}=-2$.
Since the condition (\ref{cond:log}) is $s/2 \in\Z$ in this case, 
a number of $(r,s)$ pairs can satisfy the condition for logarithms.
The condition (\ref{cond:log2}) is always satisfied, 
so the general solution for the boundary two-point functions is given in
eq. (\ref{sol:pq double}). 

Denoting the Kac field $\Phi_{1,2}$ of dimension $h_{1,2}=-1/8$ by $\mu$,
let us depict the relevant two-point functions with the boundary:
\begin{eqnarray}
&& \braket{\Phi_{r,s}(z_1,\bz_1) \mu(z_2,\bz_2)}_{boundary}
\nn
&& = (z_1 - \bz_1)^{-2h_{r,s}} (z_2 - \bz_2)^{\frac14}\;
 \xi^{\frac14\left(1+2n\right)}
\nn&&\quad\times
 \left\{
 \alpha^{(r,s)}\,
 {}_2 F_1\left(\frac12, \frac12 +|n| ; 1+|n|; \xi\right)
  + \wt \alpha^{(r,s)}\, 
 {}_2 F_1\left(\frac12+n, \frac12; 1; 1-\xi\right) 
 \right\} \,,\,\,
\end{eqnarray}
where $n=s/2 -r \in\Z$, $\alpha^{(r,s)}$ and $\wt \alpha^{(r,s)}$ are constants.
Substituting $(r,s)=(1,2)$ and $(r,s)=(2,2)$, one finds:
\begin{eqnarray}
\label{sol:mumu numu}
&& \braket{\mu(z_1,\bz_1) \mu(z_2,\bz_2)}_{boundary} 
\nn
&& = (z_1 - \bz_2)^{\frac14} (z_2 - \bz_1)^{\frac14}\;
 \xi^{\frac14} (1-\xi)^\frac14
 \left\{ \alpha^\mu\, 
 {}_2 F_1\left(\frac12, \frac12; 1; \xi\right)
       + \wt \alpha^\mu\, 
 {}_2 F_1\left(\frac12, \frac12; 1; 1-\xi\right) \right\}\,,\,\,
\nn
&& \braket{\nu(z_1,\bz_1) \mu(z_2,\bz_2)}_{boundary} 
\nn
&& = (z_1 - \bz_1)^{-\frac34} (z_2 - \bz_2)^{\frac14}\;
 \xi^{-\frac14} 
 \left\{ \alpha^\nu\, 
 {}_2 F_1\left(\frac12, \frac32; 2; \xi\right)
       + \wt \alpha^\nu\, 
 {}_2 F_1\left(-\frac12, \frac12; 1; 1-\xi\right) \right\}\,,\,\,
\end{eqnarray}
where $\nu$ denotes $\Phi_{2,2}$ of $h=3/8$, $\alpha^{\mu,\nu}$ and
$\wt \alpha^{\mu,\nu}$ are arbitrary constants.
Substituting $z_i=x_i+i\,y_i$ and $x=x_2-x_1$ into (\ref{sol:mumu numu}), one
further finds that both of them exhibit the
logarithmic singularity in the vicinity of $\xi=1$ ($i.e.$ $y_i \ll |x|$):
\begin{eqnarray}
\label{munu asym}
\braket{\mu(z_1,\bz_1) \mu(z_2,\bz_2)}_{boundary}
 &\sim& \left(y_1 y_2\right)^\frac14 \ln\left(\frac{y_1 y_2}{x^2}\right), 
\nn
\braket{\nu(z_1,\bz_1) \mu(z_2,\bz_2)}_{boundary}
 &\sim& \left(y_1\right)^{-\frac34} \left(y_2\right)^\frac14 \ln\left(\frac{y_1 y_2}{x^2}\right) .
\end{eqnarray}
Comparing (\ref{sol:mumu numu}) with the general solution (10) of \cite{KW}:
\begin{eqnarray}
&& \braket{\mu(z_1,\bz_1) \mu(z_2,\bz_2)}_{boundary} 
\nn&&
 = (z_1 - \bz_2)^{\frac14} (\bz_1 - z_2)^{\frac14} \xi^{\frac14} (1-\xi)^{\frac14}
  \left\{ A\; {}_2 F_1\left(\frac12,\frac12;1;\xi\right) + B \; {}_2 F_1\left(\frac12,\frac12;1;1-\xi\right) \right\},
\nn
\end{eqnarray}
one immediately reads:
\begin{eqnarray}
 A = \alpha^\mu {\rm ~~~and~~~} B=\wt \alpha^\mu,
\end{eqnarray}
up to phase.

From the work by Kogan and Wheater \cite{KW}, these constants are known to
be related to two-point functions of {\it logarithmic} boundary operators.
Namely, 
in the case of logarithmic theories, 
the bulk-boundary OPE relation of dimension $h$ field is to be of the
form \cite{KW,Cardy:tv}:
\begin{eqnarray}
 \Phi_h (x,y)
 = (2y)^{\Delta_d -2h} C_{\Phi\Phi}^{d} \left( d(x)+c(x) \ln(2y) \right)
 + \sum_{i} (2y)^{\Delta_i -2h} C_{\Phi\Phi}^{i} \Psi_{i}(x), 
\end{eqnarray}
where $\Psi_i(x)$'s are {\it ordinary} boundary operators of boundary scaling
dimensions $\Delta_i$, normalised as $\braket{\Psi(0)\Psi(x)}=x^{-2\Delta_i}$.
$c(x)$ and $d(x)$ are {\it logarithmic} boundary operators of dimension
$\Delta_d$, whose two-point functions are given by \cite{KW}:
\begin{eqnarray}
\label{2pt bop}
 \braket{d(0)d(x)} &=& x^{-2\Delta_d} \left(-2 \alpha_d \ln(x)+\alpha_d^\prime\right) , 
\nn
 \braket{c(0)d(x)} &=& \braket{d(0)c(x)} = x^{-2\Delta_d} \alpha_d, 
\nn
 \braket{c(0)c(x)} &=& 0 .
\end{eqnarray}
Substituting the above, $\Delta_d=0$, and $z_i=x_i+i\,y_i$ with $x_1=0$, and $x=x_2$ into the forms
$\braket{\mu(z_1,\bz_1)\mu(z_2,\bz_2)}$ and $\braket{\nu(z_1,\bz_1)\mu(z_2,\bz_2)}$, then taking the limit $y_i\ll x$,
one finds the relations between $\alpha^\mu$, $\alpha^\nu$, and $\alpha_d$ from the
asymptotic behaviours (\ref{munu asym}):
\begin{eqnarray}
\label{rel:constants}
 \alpha^\mu \sim \alpha_d \left( C_{\mu\mu}^d \right)^2 \,,\;\; 
 \alpha^\nu \sim \alpha_d \left( C_{\nu\nu}^d C_{\mu\mu}^d \right) \,,\;\; 
\end{eqnarray}
up to phase.
This result holds even if there is another logarithmic pair of $\Delta_d=1$.

Hence, with the Neumann boundary condition, the decoupling of $c(x)$,
that is $\alpha_d=0$, may occur only when 
$\alpha^\mu=\alpha^\nu=0$.
Note that, in such a case, the $\log$ term in eq. (\ref{2pt bop}) and
$\braket{\mu(z_1,\bz_1)\mu(z_2,\bz_2)}$ vanish simultaneously.

Lastly, it should be noted that, if $\wt \alpha^{\mu}=0$ $i.e.$ $B=0$, the
asymptotic behaviours of the two-point functions 
in the vicinity of $\xi=0$ match with the one-point function
$\braket{C(z,\bz)}=\braket{C(z)C(\bz)}=0$. 
We do not have this condition so that the theory may break the scaling
covariance as was discussed in \cite{KW}.

\section{Free Boson Realisation of the $SU(2)_k$ WZNW model with Boundary}
\label{sec:su2}

We begin with the quantum equivalent 
action functional of three scalar boson fields $U$, $V$, and $\Phi$ with a boundary \cite{Gerasimov:fi}:
\begin{eqnarray}
 S_{SU(2)} [U,V,\Phi] = 
 \frac{1}{8\pi} \sum_{\Phi_i=\{U,V,\Phi\}} \int_{\Sigma} d^2 z \sqrt{g}
 \left[ g^{\mu\nu} \pa_\mu \Phi_i \pa_\nu \Phi_i + 2 i \alpha_{\Phi_i} R^{(2)} \Phi_i \right] ,
\end{eqnarray}
where $\Sigma = UHP$, 
$g_{\mu\nu}=\left(\begin{array}{cc}
 0& \frac12\\
 \frac12& 0
	    \end{array}\right)$, and 
the background charges are 
$\alpha_{U}= \frac{i}{2},\, \alpha_{V}= \frac12,\, \alpha_{\Phi}= - \frac{1}{\sqrt2 \hat{q}}$. 
$\hat{q}$ gives the level $k$ of the affine Kac-Moody algebra (AKM algebra) as
$\hat{q}^2=k+2$ while the central charge is 
$c= 
\frac{3k}{k+2}$. For the theory to be analogous to the WZNW model, we
assume $k>-2$.
The EOMs imply $U=u+\bu$, $V=v+\bar v$,
$\Phi=\phi+\bphi$, and one can choose two different boundary conditions
for each boson. 

Taking only the Neumann boundary conditions into
consideration, the propagators of holomorphic fields are given by:
\label{def:prop su2}
$u(z)u(w)\sim -\ln(z-w)$, 
$v(z)v(w)\sim -\ln(z-w)$, 
$\phi(z)\phi(w)\sim -\ln(z-w)$.
Propagators of anti-holomorphic fields are given similarly, and so are
those of holomorphic and anti-holomorphic fields, while the ones
with different species such as $\braket{uv}$ vanish. 
If we introduce a notation $\phi_i = \left\{ u,v,\phi \right\}$,
the above propagators take a rather simpler form:
\begin{eqnarray}
\label{def:propagator su2k}
 \phi_i(z) \phi_j(w) \sim - \delta_{ij} \ln(z-w) . 
\end{eqnarray}
Note that all the boson propagators can be represented
by eq. (\ref{def:propagator su2k}) through Cardy's method of images.

The energy-momentum tensor can now be written in terms of $\{u,v,\phi\}$
by:
\begin{eqnarray}
   T = \sum_{\phi_i=\{u,v,\phi\}} \left( -\frac12 (\pa\phi_i)^2 + i
\alpha_{\phi_i} \pa^2 \phi_i \right) ,
\end{eqnarray}
where $\alpha_{\Phi_i}$ are relabelled by $\alpha_{\phi_i}$ for convenience.
The $SU(2)_k$ AKM currents can locally be expressed by:
\begin{eqnarray}
 J^+ &=& \frac{1}{\sqrt2}\, \pa v\, e^{-u+iv} \,,
 J^0 \,=\, \frac{i \hat{q}}{\sqrt2}\, \pa\phi + \pa u \,,
\nn
 J^- &=& \frac{1}{\sqrt2} \left[ \sqrt2 \hat{q}\, \pa\phi - i \hat{q}^2 \pa u + (1-\hat{q}^2) \pa v \right] e^{u-iv} \,.
\end{eqnarray}
They satisfy the following OPEs:
\begin{eqnarray}
  T(z) T(w) &\sim& \frac{c/2}{(z-w)^4} + \frac{2\,T(w)}{(z-w)^2} + \frac{\pa_w T(w)}{z-w} \,,
\nn
  T(z) J^a(w) &\sim& \frac{J^a(w)}{(z-w)^2} + \frac{\pa_w J^a}{z-w} {~~~for~~a=0,\pm}\,,
\nn
  J^0(z) J^\pm(w) &\sim& \frac{\pm J^\pm(w)}{z-w} \,,\;\; 
  J^\pm (z) J^\pm(w) \;\sim\; 0,
\nn
  J^0(z) J^0(w) &\sim& \frac{k/2}{(z-w)^2} \,,\;\;
  J^+(z) J^-(w) \;\sim\; \frac{k/2}{(z-w)^2} + \frac{J^0(w)}{z-w} \,,
\end{eqnarray}
up to regular terms w.r.t. $z$. Others vanish.
The anti-holomorphic currents such as $\ol T$ are given similarly and
are mapped to the lower half-plane, $\ol T(\bz)= T(z^*)$, by the method of images.

At this stage, one can follow the procedures of the $c_{p,q}$ models.
The non-chiral primaries of the $(j,m)$ highest weight representation
(hwrep) of the $SU(2)_k$ AKM algebra ($(j,\bar m)$ for the $\ol{SU(2)}_k$) are given by the chiral primaries
$\Phi_{j,m}(z)$: 
\begin{eqnarray}
\label{def:nonchiral su2}
 \Phi_{j;m,\bar m}(z,\bz) = \Phi_{j,m}(z)\,\Phi_{j,\bar m}(z^*) , 
\end{eqnarray}
whose conformal dimensions are degenerate at $h_j=\bar h_j=\frac{j(j+1)}{k+2}$.
By this mapping, their boundary two-point functions are given by chiral
four-point functions as was shown in the $c_{p,q}$ models:
\begin{eqnarray}
&&  \braket{\Phi_{j_1;m_1,\bar m_1}(z_1,\bz_1)\Phi_{j_2;m_2,\bar m_2}(z_2,\bz_2)}_{boundary} 
\nn
&& = \braket{\Phi_{j_1,m_1}(z_1)\Phi_{j_2,m_2}(z_2)\Phi_{j_2,\bar m_2}(z_2^*)\Phi_{j_1,\bar m_1}(z_1^*)}_0 . 
\end{eqnarray} 
On the other hand, one can construct the above multiplet via the vertex
operator realisation: 
\bea
  V_{j,j-m} \equiv e^{i\alpha_j \phi} e^{(j-m) (u-iv)} ,\;
 \label{def:su2dual}
  \wt V_{j,j-m} = V_{-1-j,-1-j-m} ,
\eea
with $\alpha_j = -2\alpha_{\phi} j$ and the conformal dimension $h_j = \frac12 \alpha_j (\alpha_j -
2\alpha_0) = \frac{j(j+1)}{k+2}$. $\wt V_{j,j-m}$ is the conjugate state. 
The $(j,m)$ chiral primary is realised in two ways by:
\begin{eqnarray}
 \Phi_{j,m} \sim V_{j,j-m} {\rm ~or~} \wt V_{j,j-m} .
\end{eqnarray} 

A screening charge and a screening current of dimension $(h,\bar h)=(1,0)$ is
given by\footnote{
There is a degree of freedom of having $\pa u$ in addition to $\pa v$, but
it is only equivalent up to constant factor in correlators.
}
\begin{eqnarray}
   Q_+ = \oint_{C\subset\Sigma} d\zeta\, J_+(\zeta),\;\; J_+ \equiv e^{2 i \alpha_\phi \phi} e^{-u +iv} \frac{\pa v}{i} =
 \frac{\pa v}{i} V_{-1,-1} ,
\end{eqnarray}
where $C$ is on $\Sigma$. The $(0,1)$ screening charge $\ol Q_+$ is given
similarly by the replacement of $C$ with $\ol C$ on $\Sigma^*$.
The analytically continued version of the screening charge is 
given with the closed contour $\wt C$ defined anywhere on $\C$:
\begin{eqnarray}
   \wt Q_+ = \oint_{\wt C\subset \C} d\zeta\, \wt J_+(\zeta) .
\end{eqnarray}
Note that the above screening charges are invariant under the $SU(2)_k$ transformations.

Via this realisation, one can express $N$-point functions of the
non-chiral fields,
and therefore chiral $2N$-point functions, 
by $\braket{ V_{j_1,m_1} \cdots V_{j_{2N},m_{2N}} Q_+^n}$ 
with the insertion of the background charges at infinity.
The charge neutrality (asymmetry) conditions read 
\bea
  \sum_i^{2N} \alpha_{j_i} + 2 n \alpha_\phi = 2 \alpha_{\phi},\, 
  \sum_i^{2N} m_i -n = -1 .
\eea
The first condition of the above can be rewritten by $\sum_i j_i -n = -1$.
Accordingly, the two-point function of $\tilde V_{j,j+m}$ and $V_{j,j-m}$
provides a non-vanishing two-point function, satisfying the charge conditions: 
$\braket{\tilde V_{j,j+m}(z) V_{j,j-m}(0)}_0 = z^{-2 h_j} $. This leads
to the definition of the conjugate, or dual, operators (\ref{def:su2dual}), 
and one-point functions: 
$\braket{\Phi_{j;m,m^\prime}(z,\bz)}_{boundary} \sim \frac{\delta_{m+m^\prime,\,0}}{(z-\bz)^{2h_j}}$.

\subsection{A Boundary Two-Point Function of the Doublet representations}
\label{sec:2pt su2}

Let us consider the doublet representations
$\Phi_{\pm}\equiv\Phi_{\frac12,\pm\frac12}$ of conformal dimension
$h_{1/2}$ $(\bar h_{1/2})$ $=\frac{3}{4(k+2)}$. The corresponding
chiral vertex operators are
\begin{eqnarray}
  V_+ = V_{1/2,0},\; V_- = V_{1/2,1} , \nn
  \tilde V_+ = V_{-3/2, -2} ,\; \tilde V_- = V_{-3/2, -1} .
\end{eqnarray}
Accordingly, from eq. (\ref{def:nonchiral su2}), there are four distinct non-chiral fields, $\{\Phi_{++},\Phi_{+-},\Phi_{-+},\Phi_{--}\}$.
The correlation function, analogous to $\braket{g(z,\bz)g^\dag(w,\bw)}$
of the $SU(2)_k$ WZNW model \cite{WZNW}, is 
$\braket{\Phi_{m,\bar m} \Phi_{\bar m^\prime, m^\prime}}$ 
with the indices $\bar m$ ($\bar m^\prime$) for the conjugate doublet representation. 
Contractions between fundamental representations of $SU(2)_k$ and
their conjugates result only in terms proportional to 
$\delta_{m,\bar m}\delta_{m^\prime, \bar m^\prime}$
and
$\delta_{m,\bar m^\prime}\delta_{m^\prime, \bar m}$. Reflecting the
$SU(2)_k$ invariance of the functions, the relevant correlation functions are:
\begin{eqnarray}
 \left\{
 \braket{\Phi_{- -} \Phi_{+ +}} , 
 \braket{\Phi_{+ -} \Phi_{+ -}} , 
 \braket{\Phi_{+ +} \Phi_{- -}} , 
 \braket{\Phi_{- +} \Phi_{- +}} 
 \right\} . 
\end{eqnarray}
The latter two entries are the charge conjugates of the former two.

Among them, we begin with the 
boundary two-point function $\braket{\Phi_{--}\Phi_{++}}$:
\begin{eqnarray}
\label{def:2pt--++}
 \braket{\Phi_{--}(z_1, \bz_1)\Phi_{++}(z_2, \bz_2)}_{boundary} &=& \braket{\Phi_-(z_1)\Phi_+(z_2)\Phi_+(z_2^*)\Phi_-(z_1^*)}_0
\nn
 &=& \alpha_{1}^{-++-} F_1^{-++-}(z_1,z_2) 
   + \alpha_{2}^{-++-} F_2^{-++-}(z_1,z_2),
\end{eqnarray}
where $\alpha_{1}^{-++-}$ and $\alpha_{2}^{-++-}$ are constants, and the functions
$F_i^{-++-}$ are
\begin{eqnarray}
 F_1^{-++-}(z_1,z_2) 
 &\equiv& \braket{V_-(z_1) V_+(z_2) V_+(z_2^*) \wt V_-(z_1^*)\,Q_+} \,,
\nn
 F_2^{-++-}(z_1,z_2) 
 &\equiv& \braket{V_-(z_1) V_+(z_2) V_+(z_2^*) \wt V_-(z_1^*)\,\ol Q_+} \,.
\end{eqnarray}
These turn out to be contour integrals on the branch covering of $\C$:
\begin{eqnarray}
 F_1^{-++-}(z_1,z_2) = \prod_{i<j} z_{ij}^{\alpha_i\,\alpha_j} z_{14}
  \oint_{C\subset\Sigma} d\zeta\, \frac{
  \prod_{i=1}^3 \left( z_i-\zeta\right)^{-2\alpha_\phi^2}
  \left( z_4-\zeta\right)^{3\cdot 2\alpha_\phi^2}}
  {\left( z_1-\zeta\right)\left( z_4-\zeta\right) } ,
\end{eqnarray}
where $\alpha_1=\alpha_2=\alpha_3=-\alpha_\phi$, $\alpha_4=-3\alpha_\phi$, 
and $\alpha_\phi^2= 1/2\hat{q}^2= 1/2(k+2)$. $F_2^{-++-}$ is given similarly.
Quite conveniently, the integral in the above is equivalent to that in 
eq. (\ref{eq:rs-12 vertex}) with $(p,q,r,s)=(k+2,1,0,2)$. Then, one may
follow the same reduction procedure as in the previous sections.
From the closed contours to the line integrals,
a condition $1/(k+2)\not\in\Z$ is necessary and sufficient. One may 
assume a sufficient condition $\infty>k>-1$, instead.

By the map of $\zeta\to w= \frac{z_{34}}{z_{31}}\frac{\zeta-z_1}{\zeta-z_4}$, 
$F_1^{-++-}$ amounts to:
\begin{eqnarray}
 F_1^{-++-} = e^{2\alpha_\phi^2 \pi i} (z_{13}z_{24})^{-2h_{1/2}} \xi^{\alpha_\phi^2} (1-\xi)^{\alpha_\phi^2}
  \int_0^{\xi} dw\, w^{-2\alpha_\phi^2-1} (\xi-w)^{-2\alpha_\phi^2} 
  (1-w)^{-2\alpha_\phi^2} .
\end{eqnarray}
Together with $F_2^{-++-}$, one again finds that those
integrals are the solutions of the hypergeometric differential equation (\ref{gauss eq}) with $(a,b,c)=(2 \alpha_\phi^2, 6 \alpha_\phi^2, 1+4 \alpha_\phi^2)$:
\begin{eqnarray}
\label{sol:su2-1}
 F_1^{-++-} 
  &\sim& \xi^{\frac{1}{2(k+2)}} (1-\xi)^{\frac{1}{2(k+2)}} \frac{\Gamma(-\frac{1}{k+2})\Gamma(\frac{k+1}{k+2})}{\Gamma(\frac{k}{k+2})} \,\xi^{-\frac{2}{k+2}}\, {}_2 F_1 \left(-\frac{1}{k+2},\frac{1}{k+2};\frac{k}{k+2};\xi \right) \,,\,\,
\nn
 F_2^{-++-} 
  &\sim& \xi^{\frac{1}{2(k+2)}} (1-\xi)^{\frac{1}{2(k+2)}} \frac{\Gamma(\frac{3}{k+2})\Gamma(\frac{k+1}{k+2})}{\Gamma(\frac{k+4}{k+2})}\, {}_2 F_1 \left( \frac{1}{k+2}, \frac{3}{k+2}; \frac{k+4}{k+2}; \xi\right) \,,
\end{eqnarray}
up tp phase.
Likewise, the two-point function $\braket{\Phi_{+-}\Phi_{+-}}_{boundary}$
can easily be found, with hypergeometric functions
obeying the hypergeometric differential equation of 
$(a,b,c)=(2\alpha_\phi^2, 6\alpha_\phi^2, 4\alpha_\phi^2)$.
For later convenience, we also list the function 
$\braket{\Phi_{+-}\Phi_{-+}}_{boundary}$ which in turn corresponds to the hypergeometric differential equation of 
$(a,b,c)=(1+2\alpha_\phi^2, 6\alpha_\phi^2, 1+4\alpha_\phi^2)$.
\begin{eqnarray}
\label{sol:su2-2}
  \braket{\Phi_{\epsilon_1 \epsilon_4}\Phi_{\epsilon_2 \epsilon_3}}_{boundary}
  &=& \sum_{i=1}^2 \alpha_{i}^{\epsilon_1 \epsilon_2 \epsilon_3 \epsilon_4}
      F_i^{\epsilon_1 \epsilon_2 \epsilon_3 \epsilon_4} (z_1,z_2) , 
\nn
  F_1^{++--} &\sim& \frac1k\, \xi^{\frac{1}{2(k+2)}} (1-\xi)^{\frac{1}{2(k+2)}} \frac{\Gamma(-\frac{1}{k+2})\Gamma(\frac{k+1}{k+2})}{\Gamma(\frac{k}{k+2})} \,\xi^{\frac{k}{k+2}}\, {}_2 F_1 \left(\frac{k+1}{k+2},\frac{k+3}{k+2};\frac{2k+2}{k+2};\xi \right) \,,\,\,
\nn
  F_2^{++--} &\sim& -2\, \xi^{\frac{1}{2(k+2)}} (1-\xi)^{\frac{1}{2(k+2)}} \frac{\Gamma(\frac{3}{k+2})\Gamma(\frac{k+1}{k+2})}{\Gamma(\frac{k+4}{k+2})}\, {}_2 F_1 \left( \frac{1}{k+2}, \frac{3}{k+2}; \frac{2}{k+2}; \xi\right) \,,
\nn
  F_1^{+-+-} &\sim& \xi^{\frac{1}{2(k+2)}} (1-\xi)^{\frac{1}{2(k+2)}} \frac{\Gamma(-\frac{1}{k+2})\Gamma(\frac{k+1}{k+2})}{\Gamma(\frac{k}{k+2})} \,\xi^{-\frac{2}{k+2}}\, {}_2 F_1 \left(\frac{k+1}{k+2},\frac{1}{k+2};\frac{k}{k+2};\xi \right) \,,\,\,
\nn
  F_2^{+-+-} &\sim& \xi^{\frac{1}{2(k+2)}} (1-\xi)^{\frac{1}{2(k+2)}} \frac{\Gamma(\frac{3}{k+2})\Gamma(\frac{k+1}{k+2})}{\Gamma(\frac{k+4}{k+2})}\, {}_2 F_1 \left( \frac{k+3}{k+2}, \frac{3}{k+2}; \frac{k+4}{k+2}; \xi\right) \,,
\end{eqnarray}
up to common phases for $F_1$'s and $F_2$'s respectively.
The solutions (\ref{sol:su2-1}) are the solutions shown in (2.3.14) of
\cite{Gerasimov:fi}. They and the first two $F_i$'s 
in (\ref{sol:su2-2}) are
equivalent to the solutions of the Knizhnik-Zamolodchikov equation (KZ-equation),
$\F_1^{(0)}(x)$, $\F_1^{(1)}(x)$, $\F_2^{(0)}(x)$, and $\F_2^{(1)}(x)$
in (4.10) of \cite{Knizhnik:nr} up to constant.
The general solutions are given by linear combinations of the
two corresponding hypergeometric functions for general $k>-1$.

Since those coefficients of the hypergeometric differential equation only differ by integers,
a contiguous relation of hypergeometric functions (1.4.5) in
SL\cite{slater} can be applied to $F_1^{+-+-}$. The relation reduces to:
\begin{eqnarray}
\label{id:Fiab}
  F_i^{+-+-} &=&  - F_i^{-++-} - F_i^{++--} ,
\end{eqnarray}
for $i=1,2$. From these relations, the calculations for the charge conjugate configurations were proven to be equal to the former cases [Appendix of \cite{Ishimoto:2003nb}]:
\begin{eqnarray}
 \braket{\Phi_{++}\Phi_{--}}_{boundary} \,=\, \braket{\Phi_{--}\Phi_{++}}_{boundary} \,,\;\;
 \braket{\Phi_{-+}\Phi_{-+}}_{boundary} \,=\, \braket{\Phi_{+-}\Phi_{+-}}_{boundary} \,.
\end{eqnarray}
The explicit forms of the above were shown in (\ref{sol:su2-1}, \ref{sol:su2-2}).

We are now ready to discuss the logarithms in the boundary two-point correlation functions. 
The condition for
logarithms is $4 \alpha_\phi^2\in\Z ~~i.e.$
\begin{eqnarray}
\label{cond:logsu2}
 \frac{2}{k+2}\in\Z. 
\end{eqnarray}
Therefore, some fractional levels $k<-1$ may satisfy the condition such as
$k=-4/3$ \cite{Gaberdiel:2001ny,Nichols:2003gm}. 
When $-1<k<\infty$, $k=0$ is the only choice. 
As was shown in the preceding section, the two independent
solutions coincide with each other when $c$ becomes integer while $a,b$
are not negative integers. 
In this logarithmic case, the general solution may be given by the
solutions regular at $\xi=1$, which in turn can be given by the contour
integrations with the analytically continued screening charges as in
Fig.\ref{fig:c23}. 
For general $k$:
\begin{eqnarray}
\label{sol:su2log2}
 \wt F_1^{-++-} &\sim& 
 \frac{\Gamma^2\left( 1-A \right)}{\Gamma\left(2-2A\right)} \,
 (1-\xi)^{1-2A} 
 {}_2 F_1 \left( 1+A, 1-A; 2-2A; 1-\xi \right) \,,
\nn
 \wt F_2^{-++-} &\sim& 
 \frac{\Gamma^2\left( A \right)}{\Gamma\left(2A\right)} \,
 {}_2 F_1 \left( 3A, A; 2A; 1-\xi \right) \,,
\nn
 \wt F_1^{++--} &\sim& 
 \frac{\Gamma\left(-A\right)\Gamma\left( 1-A \right)}{\Gamma\left(1-2A\right)} \,
 (1-\xi)^{-2A} 
 {}_2 F_1 \left( A, -A; 1-2A; 1-\xi \right) \,,
\nn
 \wt F_2^{++--} &\sim& 
 \frac{\Gamma\left( A \right)\Gamma\left( 1+A \right)}{\Gamma\left(1+2A\right)} \,
 {}_2 F_1 \left( 3A, A; 1+2A; 1-\xi \right) ,
\end{eqnarray}
where $A=2\alpha_\phi^2=1/(k+2)$.
$i=1$ in $F_i$'s corresponds to the interval $(z_2,z_3)$ while $i=2$ is for $(z_1,z_4)$.
Their phases and the common factor of
$(z_{13}z_{24})^{-2h_{1/2}} \xi^{\alpha_\phi^2} (1-\xi)^{\alpha_\phi^2}$
appearing in $F_i^{-++-}$ are omitted.
The logarithmic condition for the above two pairs is in common and is equivalent to the condition (\ref{cond:logsu2}):
\begin{eqnarray}
 2 A = \frac{2}{k+2} \in \Z . 
\end{eqnarray}
Therefore, the logarithmic cases for $SU(2)_k$ doublets 
necessarily contain a logarithmic solution in each limit, $\xi\to0$ or
$\xi\to 1$.

When $k=0$ or $k\to 0$, one obtains identical solutions from
(\ref{sol:su2log2}) and the general solutions are then given by: 
\begin{eqnarray}
 \braket{\Phi_{--}\Phi_{++}}_{boundary}
 &=& \alpha^{-++-} F_2^{-++-} (z_1,z_2) 
 + \wt \alpha^{-++-} \wt F_2^{-++-} (z_1,z_2)
\nn
 &=& \frac{\pi}{2} \,
  |z_1-\bz_2|^{-\frac32} \,\xi^{\frac14} (1-\xi)^{\frac14}\,
\nn&&\times
  \left\{
  \alpha^{-++-}
  {}_2 F_1 \left( \frac12, \frac32; 2; \xi \right) 
  + \wt \alpha^{-++-} \cdot 2\,
  {}_2 F_1 \left( \frac12, \frac32; 1; 1-\xi \right)
  \right\} \,,
\nn
 \braket{\Phi_{+-}\Phi_{+-}}_{boundary}
 &=& \alpha^{++--} F_2^{++--} (z_1,z_2)
 + \wt \alpha^{++--} \wt F_2^{++--} (z_1,z_2)
\nn
 &=& \frac{\pi}{2} \,
  |z_1-\bz_2|^{-\frac32} \,\xi^{\frac14} (1-\xi)^{\frac14}\,
\nn&&\times
  \left\{
  \alpha^{++--} \cdot 2\,
  {}_2 F_1 \left( \frac12, \frac32; 1; \xi \right)
  + \wt \alpha^{++--}
  {}_2 F_1 \left( \frac12, \frac32; 2; 1-\xi \right)
  \right\} \,.
\end{eqnarray}
Both solutions are symmetric to each other under the exchange of
$\xi\leftrightarrow 1-\xi$. They are logarithmic in the vicinity of $\xi=0,1$, that is, at
both the far-from-boundary limit and the near-boundary limit. 
The above expressions agree with the general solution of the chiral
four-point function in \cite{Caux:1996kq} and
the boundary two-point functions in \cite{KW} up to constant factors. 

Substituting $z_i=x_i+i y_i$
and $x=x_1-x_2$, the asymptotic behaviours at the near-boundary limit ($\xi=1$) are:
\begin{eqnarray}
\label{asym su2}
 \braket{\Phi_{--}\Phi_{++}}_{boundary}
 &\sim& \alpha^{-++-} (2 y_1)^{-\frac34} (2 y_2)^{-\frac34} \,
  \left\{- \left( \frac{4y_1 y_2}{x^2} \right)
  \ln \left(4\,\frac{y_1 y_2}{x^2}\right)\right\} \,,
\nn
 \braket{\Phi_{+-}\Phi_{+-}}_{boundary}
 &\sim& \alpha^{++--} (2 y_1)^{-\frac34} (2 y_2)^{-\frac34} \,
  \left\{2
  - \frac12 \left(\frac{4\,y_1 y_2}{x^2}\right) \ln \left(4\,\frac{y_1 y_2}{x^2}\right)\right\} \,.
\end{eqnarray}

The bulk-boundary OPE relation is to be of the form \cite{Caux:1996kq,KW}:
\begin{eqnarray}
\label{eq:b-b OPE su2}
  \Phi_{\epsilon_1 \epsilon_2}(x,y)
  =  (2y)^{-\frac34} \left\{ 
   C_{\epsilon_1 \epsilon_2}^I I_{\epsilon_1 \epsilon_2^\vee} + 
   2y \,C_{\epsilon_1 \epsilon_2}^d \, t_{\epsilon_1 \epsilon_2^\vee}^{i}
   \left( D^i(x) + C^i(x) \ln(2y) \right) + \cdots \right\} ,
\nn
\end{eqnarray}
where $I$ is the unit matrix, $t^i_{\epsilon_1 \epsilon_2^\vee}$ are the
Pauli matrices divided by two, and $\epsilon^\vee$ is the weight
conjugate to $\epsilon$. 
The two-point functions of the {\it logarithmic} boundary operators
$C^i$ and $D^i$ were given by \cite{Caux:1996kq}:
\begin{eqnarray}
 \braket{D^i(0) D^j(x)} &=& - \left( \beta + 2 \alpha_d^{su(2)} \ln(x) \right) \frac{\delta^{ij}}{x^2} \,,
\nn
 \braket{C^i(0) D^j(x)} &=& \frac{\alpha_d^{su(2)} \delta^{ij}}{x^2}\,,
\nn
 \braket{C^i(0) C^j(x)} &=& 0 .
\end{eqnarray}
We added the normalisation factor $\alpha_d^{su(2)}$ to the original expressions.
Substituting the expression (\ref{eq:b-b OPE su2}) and the above into
the forms of the boundary two-point functions, one finds the following
relations from the asymptotic forms (\ref{asym su2}) up to phase:
\begin{eqnarray}
 \alpha^{-++-} &=& 
- 
\frac12 \alpha_d^{su(2)} (C_{--}^d) (C_{++}^d) \,,
\nn
 \alpha^{++--} &=&  
\frac12 \left(C_{+-}^I \right)^2 
 \;=\; - \frac12 \alpha_d^{su(2)} \left(C_{+-}^d\right)^2 \,.
\end{eqnarray}
An interesting point is in the second relation that the normalisation
constant is related and proportional to the squared fusion coefficient
$\left(C_{+-}^I\right)^2$. This means that given fusion rules one can
fix $\alpha_d^{su(2)}$, and moreover, $\alpha^{-++-}$ and $\alpha^{++--}$. 
This feature isn't present in the
$\braket{\Phi_{r,s}\Phi_{1,2}}$ of the $c_{p,q}$ models.
Note that one can also assign different $\alpha_d^i$ instead of $\alpha_d^{su(2)}$ for different suffices
of the logarithmic pairs. In such a case, $\alpha_d^1=\alpha_d^2$ holds and
the different $\alpha_d^1$ and $\alpha_d^3$ only show up in the first
and second lines separately.

\section{Conclusions \& Remarks} 
\label{ch:conclusion}

In section \ref{sec:CG}, 
we have examined the $c_{p,q}$ models with the Neumann boundary condition in the Coulomb gas picture. 
In addition to the general $(p,q)$ cases, a class of general solutions (\ref{sol:pq double}) for the logarithmic cases were shown for the boundary two-point functions $\braket{\Phi_{r,s}\Phi_{1,2}}$ with the analytically continued screening operators. 
The cases are defined such that the 
differential equation for the function has a logarithmic solution at
$\xi=0$ and $\xi=1$. 
In its final subsection, we turned into a specific case $c=-2$ and showed
the two-point functions $\braket{\mu(z_1,\bz_1)\mu(z_2,\bz_2)}$ and
$\braket{\nu(z_1,\bz_1)\mu(z_2,\bz_2)}$, from which the relations
between the normalisation constants (\ref{rel:constants}) were deduced.

In this construction, the logarithmic condition (\ref{cond:log}) at $\xi=0$ was primarily a necessary condition. Later, it turns out to be necessary and sufficient for logarithms at $\xi=0$, because the solutions (\ref{sol:wt F1}, \ref{sol:wt F2}) are logarithmic at $\xi=0$ under the condition (\ref{assum:pq}). However, this condition is clearly not the case for logarithms at both $\xi=0$ and $\xi=1$, since (\ref{sol:pq2}, \ref{sol:wt F1}, \ref{sol:wt F2}) may all be regular at $\xi=1$. In a full theory without boundaries, if the general solution is logarithmic at $\xi=0$ while it isn't at $\xi=1$, it clearly violates monodromy invariance, or single-valuedness, and should therefore be excluded. Whereas, boundary theories do not have such a restriction so that it may have solutions which are logarithmic only at $\xi=0$. This is a novel feature from our investigation.
Note, on the other hand, that the condition (\ref{cond:log}) with (\ref{cond:log2}) are necessary and sufficient for logarithms at both points, as well as for (\ref{sol:pq2}) to be logarithmic at $\xi=1$.

Another important point on the condition (\ref{cond:log}) is that, when $q=1$, or $p \neq 1$ and $(p,q)$ coprime integers, the condition reads:
\begin{eqnarray}
  s = multiple~ of~ p .
\end{eqnarray}
With (\ref{cond:log2}), $s$ is even. Therefore $p$ is even, otherwise $s=p \times (even~integer)$.
This means that it is possible for any $c_{p,q}$ model to have such a two-point function as (\ref{sol:pq2}) and/or (\ref{sol:wt F1}, \ref{sol:wt F2}) in the presence of boundary, if one includes the $(r,s=p)$ field in the theory. This also includes $c_{p,1}$ models and the result may contain a logarithm in the vicinity of $\xi=1$. This conclusion is consistent with Flohr's conjecture and observation that the augmented conformal grid of the $c_{p,q}$ models gives the rational logarithmic CFT \cite{flohr2}.
However, it should be mentioned here that the condition (\ref{cond:log2}) singles out only $(p,q)=(2,2m+1)$ models.

In section \ref{sec:su2}, we have examined the free field realisation of the $SU(2)_k$ AKM algebras motivated by the $SU(2)_k$ WZNW action.
From the action of the three free scalar bosons with the Neumann boundary condition, we followed the calculation of \cite{Gerasimov:fi} and confirmed the general solutions (\ref{sol:su2-1}, \ref{sol:su2-2}).
With the same technique in section \ref{sec:pq}, it was found that the logarithmic cases in the doublet representation inevitably becomes logarithmic at both $\xi=0$ and $\xi=1$. 
We then found the solutions regular at $\xi=1$ and restored the general solution (\ref{sol:su2log2}) of the differential equation shown in \cite{KW}. 
Some relations between the normalisation constants and fusion rules are also listed. 
Our results also reveal that such relations may be different for different LCFTs.

Similarly, one can imagine the doublet representations in $SU(2)$ WZNW model at fractional level $k=-4/3$. But only it would not be an irreducible (or physical) highest weight representation nor an admissible representation of the theory, and therefore be excluded \cite{Gaberdiel:2001ny}. Nevertheless, it would be interesting to see whether boundary two point functions of relevant highest/lowest representations are logarithmically singular or not along our line. Related discussions on logarithmic representations in $SU(2)_{-4/3}$ WZNW model can be found in \cite{Nichols:2003gm}.

There are several points to be remarked on about our formalism.
Firstly, we have not discussed the Dirichlet boundary condition since this can
only be examined easily when $p=q$ where no screening charges are
required. 
Secondly, the puncture-type operators in \cite{Kogan:1997fd} have not been dealt with in this paper. 
If one replaces our conjugate operators with appropriate puncture-type operators and defines $\braket{:\phi\,V_{2\alpha_0}:}\sim 1$, then one finds the same result. 
Thirdly, 
the boundary operators were discussed in section \ref{sec:c=-2 bop} whereas the boundary states for LCFTs were not.
Recently, it was shown that the boundary states of CG may be realised \cite{kawaic1, Hemming:2004dm}. 
Their charge neutrality conditions are separately defined for holomorphic and anti-holomorphic sectors and would exclude our analytically continued screening operators. This leads to vanishing coefficients of $\wt \alpha$'s and therefore the restoration of the scaling covariance $B=0$, as once discussed in \cite{yi talk}. However, the Fock space structure must be considerably different from theirs and the comparison is to be studied carefully. Note that our logarithmic conditions remain intact and valid even in their way.
Another possibility of boundary crossing contours is to introduce new objects around the boundary as in \cite{schulze}. 
It may also give the logarithmic solution of eq. (\ref{gauss eq}) but our action should be modified as to be consistent with them.

There are several different models at $c=-2$. Our CG seems to realise the $c_{p,1}$ models in \cite{gab3} and the $W(2,3)$ singlet algebra in \cite{flohr1}, which does not have a nice modular property and the rationality in the same sense.
The techniques we used are quite common and, therefore, the same procedure should be applicable to many other similar models with free bosons.
In order to compare this formalism with the well-known
rational model of $c=-2$, one must have the vertex operator realisation
of the triplet algebra though \cite{kau1}. 
Recently, Nichols presented such a sketch in \cite{Nichols:2003dj} along the line of \cite{Kogan:1997fd,flohr4}.
This would be an interesting direction to study.

It would be intriguing to interpret the results in the D-brane context,
since the solutions would correspond to states on the D-branes. Once it is established, one can fix the normalisation constants $\alpha$'s in
sections \ref{sec:CG} and \ref{sec:su2} and calculate the open-closed string interactions. On the other hand, we do not know 
the physical interpretation of boundary conditions in the way that we do in the Ising model. 
These remain to be done.


\vspace{30pt}
\acknowledgments

YI is very grateful to Ian I. Kogan for his stimulating discussions with his great insight and enthusiasm. More than half of this paper was carried out by him at Oxford. 
YI is also grateful to J. F. Wheater, P. Austing, and A. Nichols for their discussions and comments.
New Century Scholarship Scheme at Oxford is acknowledged.


\end{document}